## Main Manuscript for

# Deep-Learning Based Super-Resolution Functional Ultrasound of Transient Brain-Wide Neurovascular Activity on a Microscopic Scale

Yang Cai[1], Shaoyuan Yan[1], Long Xu[1], Yanfeng Zhu[2], Bo Li[3]*, Dean Ta[1], Kailiang Xu[1]*

[1]College of Biomedical Engineering, Fudan University; Shanghai, 200438, China.

[2]Department of Neurosurgery, Huashan Hospital, Institutes for Translational Brain Research, Fudan University; Shanghai 200032, China.

[3]School of Psychology, Institute of Chinese Cultural Psychology, Shanghai Jiao Tong University, Shanghai 200030, China

*corresponding author: Kailiang Xu, Bo Li.

**Email:** xukl@fudan.edu.cn, bo-li@sjtu.edu.cn

## Abstract

Transient brain-wide neuroimaging on a microscopic scale is pivotal for brain research, yet existing imaging modalities face challenges in meeting such spatiotemporal requirements. Functional ultrasound (fUS) enables transient neurovascular imaging through red blood cell backscattering, but suffers from diffraction-limited spatial resolution. Functional ultrasound localization microscopy (fULM) has addressed this limitation by integrating ULM with fUS; but this approach requires repeated stimulation and data accumulation. Here, we introduce super-resolution functional ultrasound (SR-fUS), a deep learning-based framework that reconstructs super-resolution ULM images from contrast-free ultrafast Doppler data. By incorporating red blood cell radial gradient fluctuation priors with uncertainty-driven loss, SR-fUS enables microscopic scale hemodynamic imaging with 25-μm spatial resolution. In rat brains, SR-fUS visualized transient pain-evoked hemodynamic responses, distinguished stimulus-specific microvascular activation patterns during single-whisker stimulation, and dynamically tracked isoflurane anesthesia-induced microvascular dilation. The accuracy of SR-fUS was further preliminarily assessed through a comparative study with two-photon microscopy.

## Introduction

Transient brain-wide functional imaging is pivotal for deciphering neural mechanisms of the brain. The established modalities such as functional magnetic resonance imaging (1) and positron emission tomography (2) provide excellent depth penetration, their limited spatiotemporal resolution precludes transient microvascular-scale analysis. Optical imaging techniques (3-5) achieve microscopic resolution for neuronal activity mapping but suffer from a limited field of view, restricting its ability to resolve the large-scale information on the global neurovascular system. Through the mechanism of neurovascular coupling (6, 7), functional ultrasound (fUS) (8) has emerged as a promising alternative, providing brain-wide coverage with high temporal resolution. However, its spatial resolution (~100 μm) remains constrained by the acoustic diffraction limit, precluding visualization of functional neurovascular coupling events on a microscopic scale.

Drawing on photoactivated localization microscopy (PALM) (9), ultrasound localization microscopy (ULM) (10) has been developed by localizing and tracking microbubbles in vasculature with deep penetration, which breaks through the diffraction limit and achieves ~ 10 μm scale spatial resolution in microvascular imaging. By combining ULM and fUS, functional ULM (fULM) (11) has imaged brain-wide neurovascular activities on a microscopic scale; this technology relies on microbubbles, limiting its detection of spontaneous transient neurovascular activity. Currently, there is a lack of neuroimaging modalities capable of observing transient brain-wide functional activity at the microscopic scale.

In the study, we present a deep learning framework for super-resolution functional ultrasound imaging (SR-fUS) that breaks through the physical limits of spatiotemporal resolution in conventional ultrasound imaging. By leveraging the nonlinear mapping capabilities of deep neural networks, our method learns the mapping between contrast-free ultrasound data and super resolution ULM image, effectively bypassing the diffraction limit. We developed an uncertainty-aware reference-guided super-resolution network, which integrates high-resolution microvascular images generated from super-resolution radial fluctuations (SRRF) of red blood cells to provide structural priors, while employing uncertainty estimation to prioritize reconstruction fidelity in low signal-to-noise regions. This dual strategy achieves 4- to 7-fold improvement in spatial resolution over conventional ultrasound imaging while maintaining fUS-comparable temporal resolution, enabling functional imaging on a microscopic scale. Critically, SR-fUS unlocks transient hemodynamic detection in microvessels during single-trial stimuli, as validated in both whisker stimulation and nociceptive pain paradigms. Furthermore, comparative studies with two-photon microscopy demonstrate the consistent functional hyperemia profiles across identical cortical microvessels in awake rats, establishing the reliability of SR-fUS.

## Results

**SR-fUS based on deep learning with spatiotemporal priors and uncertainty-driven loss**

Our objective is to achieve SR-fUS through an advanced deep learning mapping strategy that transforms low-resolution ultrafast Doppler images into super-resolution ULM images. ULM images can be regarded as the standard vascular distribution, whereas ultrafast Doppler images are low-resolution representations of blood flow signals. The SR-fUS can actually be formulated as an inherently ill-posed image restoration problem (12), meaning that a given ultrafast Doppler input corresponds to infinitely many possible ULM reconstructions in a high-dimensional space, resulting in reconstruction errors and artifacts. This challenge is further compounded by the low signal-to-noise ratio (SNR) of contrast-free ultrafast Doppler images. To address this limitation, it is essential to introduce constraints or priors. Red blood cells are substantially smaller than the acoustic wavelength, they act as Rayleigh scatterers. In ultrasound imaging, the coherent accumulation of their scatterings within the resolution cell generates speckle patterns. These speckle patterns exhibit intensity peak statistically centered at the vascular center (13, 14). The gradient field within a neighborhood surrounding each peak demonstrates radial-like convergence toward the centroid (Fig. S1). Harnessing this gradient convergence enables enhanced blood flow signal detection and delivers high-resolution microvascular structural information. Furthermore, blood flow data exhibit correlated temporal fluctuations distinct from noise, enabling temporal analysis to enhance SNR in contrast-free ultrasound imaging. Given that red blood cells and microbubbles share similar flowing pattern and data characteristics in the blood flow, we hypothesize that cross-domain mapping can transform red blood cell data into microbubble-equivalent representations, thereby enabling contrast-free super-resolution imaging by leveraging knowledge distilled from ULM data. Deep learning (15, 16) provides an ideal framework for this transformation, which has already demonstrated significant potential in super-resolution imaging (17-22). Inspired by these advances, we developed SR-fUS, a deep learning approach that reconstructs ultrafast Doppler images into super-resolution Doppler images to detect functional hyperemia changes with high spatiotemporal resolution.

Figure 1A illustrates the SR-fUS data processing pipeline. In conventional ultrasound imaging, blood flow images (ultrafast Doppler) are generated from sequences of B-mode images acquired at ultrafast frame rates (1 kHz), exhibiting low spatial resolution. SR-fUS employs deep learning to reconstruct ultrafast Doppler images into super-resolution Doppler images, using ULM images (each generated from 300 s of microbubble data) as training supervision targets. Cerebral blood volume (CBV) changes can then be extracted from the super-resolution Doppler sequence for functional brain activity evaluation. This framework achieves spatial resolution comparable to ULM (~25 μm) while maintaining the native temporal resolution of ultrafast Doppler, enabling SR-fUS imaging. Comparative studies with two-photon microscopy validated the reliability of SR-fUS, demonstrating strong agreement in microvascular structure and functional dynamics (Fig. 1B).

We designed an uncertainty-driven reference-guided super-resolution network (URSR Net) with the specific architecture shown in Fig. 1C. We employed the residual-in-residual dense block net (RRDBNet) (22) as the backbone network to learn the complex mapping from low-resolution ultrasound data to super-resolution ULM images, leveraging its established efficacy in image super-resolution tasks. For the network input, while ultrafast Doppler images provide high temporal resolution and blood volume information, they alone are insufficient for accurate microvascular reconstruction (Fig. S2A). To compensate for the loss of spatial information in ultrafast Doppler images, we incorporated high-resolution reference images generated by super-resolution radial fluctuations (SRRF) (23) as an additional input. SRRF enhances spatial resolution in diffraction-limited images by exploiting radial fluctuations across image sequences, which has already been applied in optical and ultrasound imaging fields (24-27). This integration enables reconstruction of fine microvascular details when combining ultrafast Doppler and SRRF inputs (Fig. S2). In our study, each ultrafast Doppler image and each SRRF image is generated from 200 frames of spatiotemporal contrast-free ultrasound data. As shown in Fig. 1C, these inputs undergo fusion via a feature fusion module before being processed by RRDBNet for feature extraction. The output layer reconstructs feature maps into super-resolution Doppler images, with an additional branch estimating pixel reconstruction variance (see Supplementary materials for details). Higher variance indicates greater reconstruction uncertainty and identifies artifact-prone regions. Consequently, we incorporated an uncertainty attention mechanism into the loss function to prioritize these challenging areas. The network was trained by optimizing a comprehensive uncertainty-driven loss function using ULM images as ground truth.

**Super-resolution Doppler achieves microvascular imaging of the rat brain**

The performance of the proposed method was testified on an *in vivo* rat brain dataset. The training dataset comprised 4 860 pairs of ultrafast Doppler images and SRRF images with corresponding ULM images, collected from 90 planes (three planes per rat brain) across 30 rat brains. For each plane, three pairs of SRRF and ultrafast Doppler images are generated. Then the dataset was augmented 18-fold through rotation, translation and Fourier-based directional filtering (28). Finally, we trained our deep neural network using ULM images (each frame was reconstructed using 300s of ultrasound data) as targets.

The trained model was then applied to a rat brain dataset which was never used for training. Figure. 2A presents a super-resolution Doppler image, while Figs. 2B and 2C show the corresponding ULM and ultrafast Doppler images, respectively. To obtain the directional figures, Fourier-based directional filtering was used to separate upward and downward data. Due to the diffraction limit and lack of contrast enhancement, ultrafast Doppler images exhibited poor spatial resolution and low SNR. Conversely, super-resolution Doppler and ULM exhibited clear and continuous microvasculature at high spatial resolution. Iso-frequency plots of ULM, super-resolution Doppler and ultrafast Doppler images are shown in Fig. 2F. Based on the half wavelength amplitude cutoff, the global spatial resolution of super-resolution Doppler was 3.5 times better than ultrafast Doppler and worse than half that of ULM. Cross-sectional vessel profiles (Fig. 2E) show our method reduced the full width at half maximum (FWHM) of microvessels from 57 μm (ultrafast Doppler) to 9 μm, resolving microvessels spaced 20 μm apart. These results indicate a 4- to 7-fold spatial resolution improvement in super-resolution Doppler compared to ultrafast Doppler.

As any image super-resolution reconstruction method, SR-fUS is susceptible to reconstruction artifacts. As shown in the left part of Fig. 2H, low SNR in dense microvascular regions produced artifacts manifesting as discontinuous vessels. When corresponding ULM images are available, reconstruction performance can be evaluated using Haar wavelet kernel analysis (29). Figure 2G presents a Haar wavelet confidence map where red regions indicate low-confidence structures. However, this method is inapplicable to contrast-free data where ULM images are unavailable. As previously mentioned, uncertainty estimation can provide an alternative assessment of reconstruction fidelity. Our network outputs a pixel variance map alongside a reconstructed image (Fig. 2H), with high-variance regions (red) indicating reconstruction uncertainty and potential artifacts. Comparison between Haar confidence maps and uncertainty maps reveals overlapping error-prone regions, validating the uncertainty estimation's effectiveness. Moreover, incorporating uncertainty information can enhance reconstruction performance. By including an uncertainty-aware mechanism in the loss function (Method URSR Net), the network is capable of better focusing on microvessels. Compared to the result from the network trained without incorporating uncertainty information (the left part of Fig. 2I), the network trained with the uncertainty-aware loss function (the right part of Fig. 2I) exhibited more continuous and smoother microvasculature, with significantly reduced artifacts in challenging regions (white dotted box in Fig. 2I). The corresponding deviation maps (Fig. 2J) comparison with ULM also indicated that the uncertainty-aware mechanism effectively reduced reconstruction errors.

We next evaluated the model's temporal resolution, which is crucial for detecting blood flow changes. In SR-fUS, temporal resolution depends on input data volume. The bottom of Fig. 2A and Fig. 2B demonstrate frame count impacts on reconstruction quality. In ULM, vascular reconstruction relies on the accumulation of microbubble trajectories. Consequently, at low data frame counts (< 5000 frames), ULM lacks sufficient microbubble trajectories to reconstruct the vasculature, resulting in poor image quality, which manifests as disconnected and missing vessels (white arrows in Fig. 2B). As data acquisition time increases (> 25 000 frames), microbubbles gradually perfuse the entire microvascular network, leading to improved image quality and vascular saturation. By contrast, the signals from red blood cells in blood flow are much denser, providing sufficient information to reconstruct the vasculature even with data acquired over a short period. Therefore, SR-fUS, which utilizes red blood cells for vascular reconstruction, can generate a complete vascular network even with a low frame count (<1000 frames). The overall vascular saturation remains above 80\%, and the multi scale structural similarity index measure (MS-SSIM) with perfect ULM images (reconstructed from 60 000 frames) exceeds 0.7. Overall, compared to ULM, super-resolution Doppler improves temporal resolution while maintaining a high reconstruction quality.

**SR-fUS reveals single-microvessel-scale functional hyperemic changes in a rat brain during whisker stimulation**

Leveraging neurovascular coupling, fUS reveals neural activity through hemodynamic monitoring. Therefore, accurate hemodynamic representation is essential for functional imaging applications. While the deep learning model employs a nonlinear cross-modality mapping, we validated that the reconstructed SR-fUS signals maintain a linear correspondence with true hemodynamic dynamics. To verify this, we monitored pulsatile blood flow in a rat brain model. The results demonstrated SR-fUS accurately captures pulsatile CBV dynamics across cortical, deep brain, and whole-brain regions (Fig. S3 and Movie S2). The temporal curves exhibit >0.9 correlation with ultrafast Doppler, confirming the ability of SR-fUS to preserve hemodynamics in ultrasound data.

This hemodynamic accuracy enables robust functional studies. We validated the proposed method on a whisker-stimulated rat brain dataset. Figure 3A and Fig. 3B present activation maps derived from SR-fUS and fUS, both showing strong activation in the vessels within the barrel field of the primary somatosensory cortex (S1BF). The activated regions detected by SR-fUS corresponded closely with the areas identified by the fUS, indicating SR-fUS can detect task-evoked cortical activation with sufficient sensitivity while providing significantly enhanced spatial resolution.

Enhanced spatial resolution enables SR-fUS to distinguish different vascular compartments, such as pial vessels and small vessels (Fig. 3C), which is challenging in conventional fUS. Figure 3B demonstrates significantly greater relative CBV changes during whisker stimulation in small vessels versus pial vessels (38% ± 7% vs. 12% ± 2%, mean ± s.e.m.), indicating their predominant role in neurovascular coupling, which is consistent with fULM findings \cite{fULM}. In contrast, conventional fUS could only detect minor differences (34% ± 8% vs. 32% ± 7%, mean ± s.e.m.). In addition, SR-fUS can allow the scale of detecting activation to be improved from vascular regions to a single microvessel (indicated by the red dashed line in the Fig. 3D). Single microvessel involvement during functional hyperemia can also be quantified. In a representative microvessel within the activated region, we quantified increases in diameter and CBV (Fig. 3D and Movie. S1). Both fluctuated regularly with the stimulus, whereas no variation was observed in the controls. Then diameter analysis was further conducted on eight different microvascular segments, and consistent results were obtained (Fig. S4). These results demonstrate the capacity of SR-fUS to resolve microscopic functional hyperemia in rat brain microvessels during whisker stimulation.

Next, to verify the ability of SR-fUS to measure independent information at the microscopic scale (i.e., functional resolution), we designed a single-whisker stimulation experiment. Two individual whiskers (C3 and D2) were stimulated separately to activate two adjacent barrels in cortex. Figures 3E and 3F show the activation maps obtained with SR-fUS and conventional fUS under the two stimulation conditions. Both modalities detected regional responses in adjacent barrel regions, indicating they are able to capture single whisker-evoked hemodynamic activation. Compared to fUS, SR-fUS shows less overlap between the activation regions corresponding to the two stimuli. This overlap may result from a spread of the hemodynamic signal (30, 31), and the reduced overlap may indicate that SR-fUS has higher functional resolution. To quantify functional resolution, we defined a pixel-resolvability index (32) and focused on the transition region between two barrels. The results show that SR-fUS can discriminate the responsiveness of neighboring pixels with a functional resolution of 37.5 μm, which is significantly better than the 200-μm resolution of fUS (Figs. 3E and 3F). Furthermore, SR-fUS can distinguish the specific responses of pixels within 25 μm to two different whisker stimuli. Figures 3G and 3H show the CBV changes detected by SR-fUS and conventional fUS during stimulation of two individual whiskers. SR-fUS identified a microvessel located near the boundary of the C3-associated barrel field that exhibited a stimulus-specific response, with a larger CBV increase during C3 stimulation than during D2 stimulation (4% vs. 1.4%). In contrast, conventional fUS failed to detect this differential response, showing similar CBV changes under the two stimulation conditions (4% vs. 3%). These results indicate that SR-fUS can not only improve spatial resolution but also enable the detection of functional hemodynamics with superior functional resolution.

**SR-fUS reveals an increase in microvascular diameter induced by isoflurane anesthesia**

One major advantage of SR-fUS over ULM and fUS is its ability to detect dynamic changes of microvascular diameter. To validate this capability, an isoflurane-induced vasodilation experiment was conducted. The rat brain was scanned during isoflurane anesthesia. Initially, the anesthesia was maintained with 1% isoflurane and after 200 s, the concentration was increased to 2% isoflurane. The entire data acquisition lasted 500 s. For comparison, ULM imaging was performed after the anesthesia level had stabilized at 1% isoflurane and 2% isoflurane.

Figure 4A compares SR-fUS images obtained at two different levels of anesthesia, showing a global increase in blood flow intensity as the isoflurane concentration rises. To quantify isoflurane-induced vasodilation, the diameters of eight microvessels (indicated by the white lines in Fig. 4A) across the brain were measured. The mean vessel diameter increased from 41 μm under 1% isoflurane anesthesia to 45 μm under 2% isoflurane anesthesia. The diameters measured by SR-fUS were consistent with the corresponding ULM measurements (Fig. 4B), with a mean error of 3.8 μm, demonstrating the accuracy of SR-fUS for microvascular diameter quantification.

Furthermore, SR-fUS allows for the investigation of specific responses to isoflurane anesthesia in arteries and veins. In the cerebral cortex, vessel type can be inferred from flow direction: downward flow primarily corresponds to cortical arteries, whereas upward flow corresponds to cortical veins. Accordingly, Fourier-based directional filtering was used to separate upward- and downward-flow components for arterio-venous analysis. Statistical analysis showed that increasing the isoflurane concentration resulted in a significant dilation of arterial diameter measured by SR-fUS (37 μm at 1% isoflurane vs. 44 μm at 2% isoflurane). In contrast, venous diameter did not exhibit a significant change. This arterio-venous difference is consistent with ULM measurements and previous studies (33, 34). Importantly, unlike ULM, which provides static microvascular diameter measurements, SR-fUS enables continuous measurements of diameter dynamics. As shown in Fig. 4E, after increasing the isoflurane concentration at 200 s, SR-fUS captured gradual dilation of both arteries and veins. The arterial diameter rapidly increased from 40 μm to 44 μm within 50 s, whereas venous dilation was slower and smaller, increasing by only two μm over 150 s. These results demonstrate the ability of SR-fUS to resolve vessel-type-specific microvascular diameter dynamics.

Beyond the cortex, SR-fUS enables brain-wide microvascular analysis. SR-fUS images of the ROI regions in the cortex, hippocampal formation and thalamus are shown in Fig. 4F. Three microvessels were selected from each region, and their relative diameter changes are quantified (Fig. 4G). The results show that vasodilation is more pronounced in the hippocampal formation and thalamus than in the cortex, with mean diameter increases of 15% and 16%, respectively, compared with 6% in the cortex. These findings are consistent with previous ULM and fMRI studies (35, 36). Furthermore, SR-fUS can capture the dynamic evolution of vasodilation in these regions (Fig. 4H), demonstrating its potential for investigating anesthesia-induced cerebrovascular regulation and brain-wide microcirculatory dynamics.

**Reliability of SR-fUS verified by comparison with two-photon microscopy**

To validate the ability of SR-fUS to resolve microvascular structures and functional dynamics, we conducted cross-modal comparisons with two-photon microscopy (37). Awake, craniotomized rats underwent SR-fUS imaging during whisker stimulation (Fig. 5A). Following functional ultrasound acquisition, a cranial window was implanted over the somatosensory cortex. After 24-hour of recovery, two-photon microscopy imaging of the same vascular network was performed.
Figure 5B displays two-photon microscopy imaging of cortical vasculature in 3D and xz-plane cross-sections. To spatially align the two-photon microscopy volumetric z-stack with the ultrasound coronal plane (xz-plane cross-section), the approximate y-axis position based on stereotaxic coordinates recorded during ultrasound imaging was first localized. Then the y-axis position within the two-photon microscopy volume was systematically shifted (step size: 10 μm) and MS-SSIM between two-photon microscopy's xz slices and SR-fUS images was computed to determine the optimal alignment. As shown at the bottom of Fig. 5B, SR-fUS reconstructed cortical microvasculature with high fidelity to two-photon microscopy (yellow dotted line in the lower right of Fig. 5B), resolving penetrating vessels spaced <40 μm apart.
Both modalities performed functional imaging during whisker stimulation. Figure. 5C shows representative images at resting state and stimulation timepoints, with functional hyperemia in microvessels (white arrowheads) clearly resolved by SR-fUS. To quantify these effects observed on representative vessels, we analyzed cross-sections of microvessels at n = 10 locations. Both modalities captured stimulus-evoked vasodilation in penetrating arterioles with high temporal correlation (Pearson's r = 0.86 in SR-fUS vs. r = 0.91 in two-photon microscopy; Fig. 5D). The quantitative SR-fUS estimated diameter ($D_{mean}$ ~ 27 ± 4 μm) of microvessels at rest, as well as their relative changes during activation (ΔD = +13 ± 3%), were consistent with measurements made using two-photon microscopy in the cortex (respectively $D_{mean}$ ~ 38 ± 3 μm and ΔD = +7 ± 3%). The above results indicate that SR-fUS is highly reliable for detecting functional hemodynamics of microvessels.

### Single-trial detection of pain-evoked microvascular dynamics via SR-fUS

To validate the ability of SR-fUS to detect transient functional activity in a single trial, we applied it to rat pain models, where pain stimulation was applied by pinching the tail (detailed in Methods). CBV correlation maps revealed brain regions activated during the post-stimulus period (Fig. 6A), with both fUS and SR-fUS indicating positive stimulation correlations in the retrosplenial granular cortex (RSGc) and mediodorsal thalamus (MD), but negative correlations in the hippocampus (HPC). CBV sequence maps corroborated these findings (Fig. S5 and Movie S4), showing rapid CBV increases in RSGc but gradual decreases in HPC following painful stimuli (Fig. 6D). These findings align with reports implicating RSGc and HPC in pain processing and modulation (38, 39).
SR-fUS also provided brain functional connectivity (FC) information consistent with fUS (Fig. 6B). Based on FC information, we performed seed-based analysis of the FC alterations with RSGc as the seed region, which is most relevant to pain-induced arousal as indicated by our previous observations. The analysis showed that after pain stimulation, both SR-fUS and fUS showed weakened FC between the RSGc and MD, as well as between the RSGc and posterior thalamic nuclear group (Fig. 6C and Fig. S6A). Whole-brain FC analysis also showed similar results, revealing a decrease in whole-brain FC after pain stimulation (Fig. S6B).

## Discussion

In the present study, an uncertainty-driven reference-guided super-resolution network is developed for SR-fUS. Based on the similar flow characteristics of red blood cells and microbubbles, our method learns the mapping between contrast-free ultrasound data and super resolution ULM image, overcoming the physical limits of spatiotemporal resolution in conventional ultrasound blood flow imaging. Through incorporating radial fluctuation information of red blood cells and uncertainty-driven loss, SR-fUS enhances the spatial resolution of fUS by a factor of 4-7. Our method paves the way for high spatiotemporal resolution fUS towards transient brain-wide neurovascular imaging on a microscopic scale.

SR-fUS enhances super-resolution reconstruction performance by incorporating structural priors derived from SRRF analysis. The low SNR of contrast-free ultrasound data and the significant resolution disparity between ultrafast Doppler images and ULM ground truth complicate learning the mapping between ultrafast Doppler images and ULM ground truth. When SR-fUS process ultrafast Doppler inputs without structural priors, it exhibited suboptimal reconstruction performance, manifesting as aberrant vascular curvatures and spurious vessel coalescence (Fig. S2B). Compared with ultrafast Doppler, SRRF resolves finer microvascular structures by exploiting radial fluctuations in red blood cell signals (Fig. S2D). Therefore, integrating SRRF-generated priors improves reconstruction fidelity, yielding accurate vascular morphology with reduced artifactual coalescence.

Uncertainty attention mechanism further enhances reconstruction fidelity while providing quantitative reliability assessment. A critical limitation in deep learning-based super-resolution imaging is unreliable predictions, where conventional deep learning methods produce overconfident outputs that manifest as artifacts—particularly in low-SNR regions of ultrafast Doppler data. To mitigate this, we combine neural networks and Bayesian learning, jointly predicting both the super-resolution Doppler (mean) and its uncertainty (variance). This predicted uncertainty map identifies regions with high probability of reconstruction errors, delivering essential reliability metrics for super-resolution imaging. Furthermore, by incorporating uncertainty estimates into the model training process through pixel-wise attention weighting of the loss function, we achieve additional reconstruction performance gains.

The spatial resolution of super-resolution Doppler was firstly evaluated in in vivo rat brains. We compared the global spatial resolution performance of the ultrafast Doppler, ULM and super-resolution Doppler in the Fourier domain (Fig. 2F) and then measured FWHM of vessels in ROI (Fig. 2E). The results indicate that the proposed method can achieve a spatial resolution improvement of about 4 to 7 times compared to ultrafast Doppler. For the temporal resolution, we can achieve a temporal resolution of 0.2 s.

However, a key question for super-resolution functional imaging is whether improved spatial resolution translates into improved functional resolution. Spatial resolution describes the ability to separate adjacent vascular structures, whereas functional resolution reflects the ability to distinguish neighboring pixels or microvessels that carry independent hemodynamic information. To examine this question, we further analyzed microvascular functional responses within barrel fields evoked by individual whisker stimulation, with particular focus on response differences between neighboring microvessels located near the boundary of adjacent barrel fields. Although both SR-fUS and conventional fUS detected C3- and D2-evoked activation in adjacent barrel regions, conventional fUS showed broader and more overlapping activation patterns. This overlap is likely caused by spatial mixing and partial-volume effects within the diffraction-limited point-spread function of conventional Doppler imaging. By mapping diffraction-limited Doppler signals onto microvascular-scale flow distributions, SR-fUS reduces this spatial mixing to improve functional resolution. To quantify the functional resolution, we defined a discriminability index between pixels. The results show that SR-fUS can discriminate the responsiveness of neighboring pixels with a functional resolution of 37.5 μm, which is significantly better than the 200-μm resolution of fUS. And SR-fUS can distinguish the specific responses of pixels within 25 μm to two different whisker stimuli. These results indicate that SR-fUS resolves stimulus-specific hemodynamic differences at the microvascular scale, demonstrating that its improved spatial resolution enhances not only vascular visualization but also the extraction of locally independent functional information.

Another important advantage of SR-fUS is its ability to track dynamic changes in microvascular diameter. ULM provides super-resolution vascular maps and accurate static vessel morphology, but it requires microbubble injection and typically accumulates data over a relatively long acquisition period. Therefore, ULM is less suited for continuous monitoring of fast or evolving diameter changes. Conventional fUS, on the other hand, provides high temporal resolution but lacks sufficient spatial resolution to reliably quantify microvessel diameters. The isoflurane experiment demonstrates that SR-fUS bridges this gap. SR-fUS diameter measurements were consistent with ULM, with a mean error of 3.8 μm, and detected an increase in mean microvessel diameter from 41 μm under 1\% isoflurane to 45 μm under 2% isoflurane. These results validate the accuracy of SR-fUS for microvascular diameter quantification and demonstrate its ability to monitor dynamic vasodilation.
What's more, SR-fUS enables to detect transient neural activity on a microscopic scale. In the functional experiment during whisker stimulations, SR-fUS can not only detect activation in relevant brain regions with high spatial resolution, but also provide quantitative parameters at single-vessel scale. Critically, SR-fUS permits reliable detection of single-trial neural responses across the microvascular hierarchy—a capability that remains challenging for fULM in microvasculature, as sparse microbubble sampling limits sensitivity.
To validate reliability of SR-fUS to resolve microvascular structures and functional dynamics, a comparative study with two-photon microscopy was performed. The microvascular results under the same facet corresponded well and the functional response of the microvessels was consistent, indicating the reliability of the method.
Despite the impressive performance of SR-fUS, its further improvement can be envisaged. Currently, this technology is not yet capable of handle blood flow velocity. To infer blood flow velocity, it is necessary to model spatiotemporal ultrasound data, which significantly increases computational complexity and modeling difficulty. Besides extracting spatial features required for super-resolution reconstruction, the network must capture temporal information between frames, which is challenging for the current model. In future work, we will explore the potential of deep learning models capable of handle spatiotemporal sequences, such as RNNs and transformers.
In summary, SR-fUS achieves contrast-free super-resolution ultrasound imaging, providing a tool for transient functional neuroactivity studies. This approach overcomes the inherent limitations of both red blood cell-based ultrasound imaging (low SNR) and microbubble-based ultrasound imaging (poor temporal resolution) through deep learning, enabling detection of single-trial functional responses in microvasculature at approximately 20-μm spatial resolution and fUS-comparable temporal resolution.

## Materials and Methods

### Uncertainty-driven reference-guided super-resolution network

URSR Net reconstructs SR vascular images from paired ultrafast Doppler and SRRF inputs using an RRDBNet-based backbone. The network incorporates feature fusion module and dual-output branches for image super-resolution reconstruction and uncertainty estimation (see Supplementary materials for details).

### Evaluation metrics

**Measurement of spatial resolution.** To evaluate the global spatial resolution of the SR reconstructed image, we compared the resolution performance of ultrafast Doppler, SRRF, ULM and the SR reconstructed image in the Fourier domain (40). The iso-frequency curves were plotted by calculating the mean value of the frequency component with the same spatial frequency. The amplitude of ultrafast Doppler at half wavelength spatial frequency was marked and the same amplitude was used to find the corresponding spatial frequency of the SR reconstructed image to determine the spatial resolution (41).
**Resolution of functional ultrasound imaging.** The method used here for measurement of functional resolution is similar to those described in (32) (see Supplementary materials for details).

**Measurement of vessel diameter of fULM.** For the measurement of vessel diameter, we have adopted the same standard as fULM (11): The diameter of the vessel is defined as the width of the vessel at a threshold set by the half maximum of the baseline profile.

### Data acquisition

The proposed deep learning-based SR reconstruction method was trained on in vivo rat brain data from 30 male Sprague–Dawley rats. Before ultrasound imaging, rats underwent craniotomy under isoflurane anesthesia. For each rat, the L22-14vX probe was positioned in the coronal plane, and three brain planes were acquired.
Ultrasound data were collected using a high-frequency linear-array transducer (L22-14vX, Verasonics Inc., Kirkland, WA) connected to a Vantage 256 system. Imaging was performed with 18-angle compounded plane waves at a transmission frequency of 15.625 MHz, PRF of 18 kHz, and post-compounding frame rate of 1000 Hz. After contrast-free ultrasound acquisition, ULM imaging was performed in the same plane to provide ground-truth SR images.

### Functional rat study

To evoke the somatosensory barrel cortex, whiskers were deflected at 3 Hz using a motorized whisker stimulator. The stimulation pattern consisted of a 30-s resting state followed by 30-s stimulation and 30-s recovery repeated 5 times, for a total of 330 s.
To allow single-whisker stimulation, all whiskers but the two individual whiskers (C3 and D2) on the left snout were trimmed before imaging. Each whisker stimulation repetition consisted of a 10-s rest period, a 10-s stimulation and a 10-s post-stimulation repeated 20 times, for a total 600 s.
In single-trial pain stimulation experiments, we applied pain stimuli by clamping the rat' tail. The stimuli were administered for 5 s after 30 s of resting under anesthesia. A total of 68 s of ultrasound data were acquired before and after the stimuli.

### Acknowledgments

The authors thank Dr. Lingxuan Zhou for helpful discussions.

### Funding

This work was supported by National Natural Science Foundation of China [No. T2525028] and Shanghai Pilot Program for Basic Research - Fudan University [No. 21TQ1400100 (25TQ002)].

### Author Contributions

Yang Cai: Conceptualization (Lead); Data curation (Equal); Methodology (Lead); Visualization (Equal); Writing - original draft (Lead).

Shaoyuan Yan: Methodology (Supporting).

Long Xu: Data curation (Equal).

Yanfeng Zhu: Data curation (Equal).

Bo Li: Supervision (Equal).

Dean Ta: Resources (Equal).

Kailiang Xu: Project administration (Equal); Resources (Equal); Software (Equal); Supervision (Equal); Writing - review & editing (Equal).

## Figures

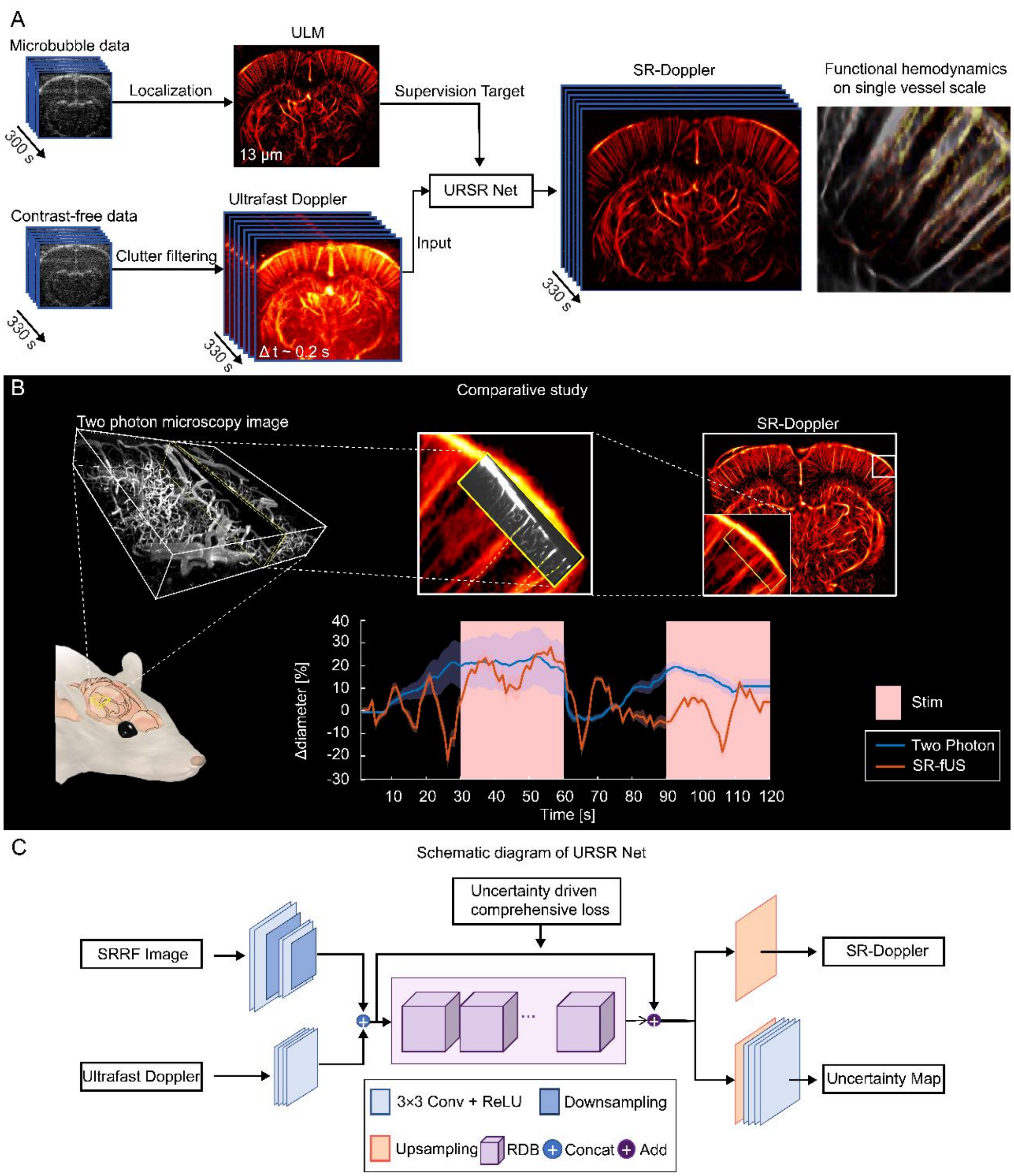

**Figure 1. Deep learning-based Super-resolution function Ultrasound (SR-fUS) framework.** (**A**) SR-fUS framework. ULM data (generated from 300s of microbubble data with a spatial resolution of 13 µm) are used to learn the mapping between contrast-free ultrafast Doppler images and their super-resolved counterparts, enabling super-resolution Doppler reconstruction with 25-µm spatial and 0.2-s temporal resolution. This super-resolution Doppler data can then be leveraged for SR-fUS. (**B**) Comparative study between the SR-fUS and two-photon microscopy. The two modalities were used to image the same cortical region in a rat brain during whisker stimulation. The major blood vessels are well-aligned between the super-resolution Doppler and two-photon microscopy images, and the functional diameter changes of these vessels also show good agreement. (**C**) Schematic diagram of the Uncertainty-aware Reference-guided Super-

Resolution network (URSR Net). Ultrafast Doppler images and SRRF images generated from spatiotemporal contrast-free ultrasound data are used as network inputs to output SR microvascular images and corresponding uncertainty maps. The deep neural network was trained by optimizing a comprehensive loss function of the output with the ULM images.

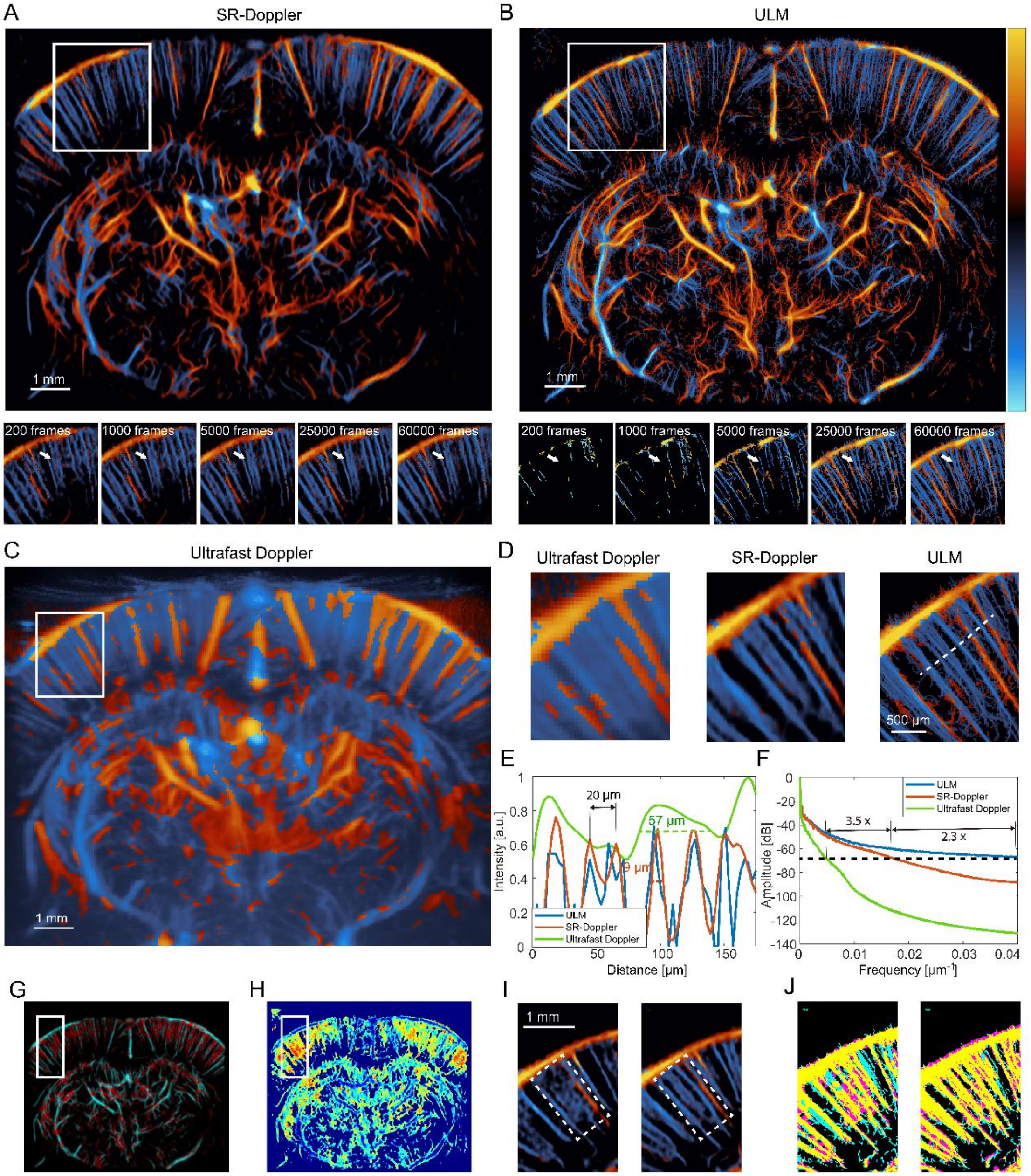

**Figure 2. Results of super-resolution Doppler in *in vivo* rat brain.** (**A** and **B**) Comparison of super-resolution Doppler and ULM in in vivo rat brain ultrasound data. Direction flow maps were reconstructed from 60,000 frames (a total of 300 s of acquisition) of contrast-free ultrasound data (for super-resolution Doppler) or CEUS data (for ULM). The bottom of the figure illustrates the impact of increased frame count on image quality. (**C**) Ultrafast Doppler image of the same rat brain. (**D**) Magnified view of ultrafast Doppler, super-resolution Doppler and ULM images marked by the white ROI in A, B and C. (**E**) Intensity profile of vessels along the white dashed line in D. (**F**) The iso-frequency plots of ultrafast Doppler image, super-resolution Doppler image and ULM image. Spatial resolution is determined using the amplitude of the ultrafast Doppler at half wavelength spatial frequency as the threshold (indicated by the black dashed line), revealing that the spatial resolution of super-resolution Doppler is 3.5 times better than that of ultrafast Doppler and 2.3 times worse than that of ULM. (**G** and **H**) The confidence map of the super-resolution

Doppler measured by haar wavelet kernel analysis (left) and the corresponding uncertainty map (right), where blue indicates well-reconstructed vessels and red indicates poor reconstruction. (**I**) Magnified view of the super-resolution Doppler images with (left) and without uncertainty attention mechanism (right), marked by the white ROI in H. Both images were reconstructed from 1,000 frames (a total of 5 s of acquisition) of contrast-free ultrasound data. (**J**) Deviation maps corresponding to H. Areas of super-resolution Doppler images agreement with ULM images are displayed in yellow whereas magenta and cyan indicate structure only present in SR-fUS reconstructed images and ULM images, respectively.

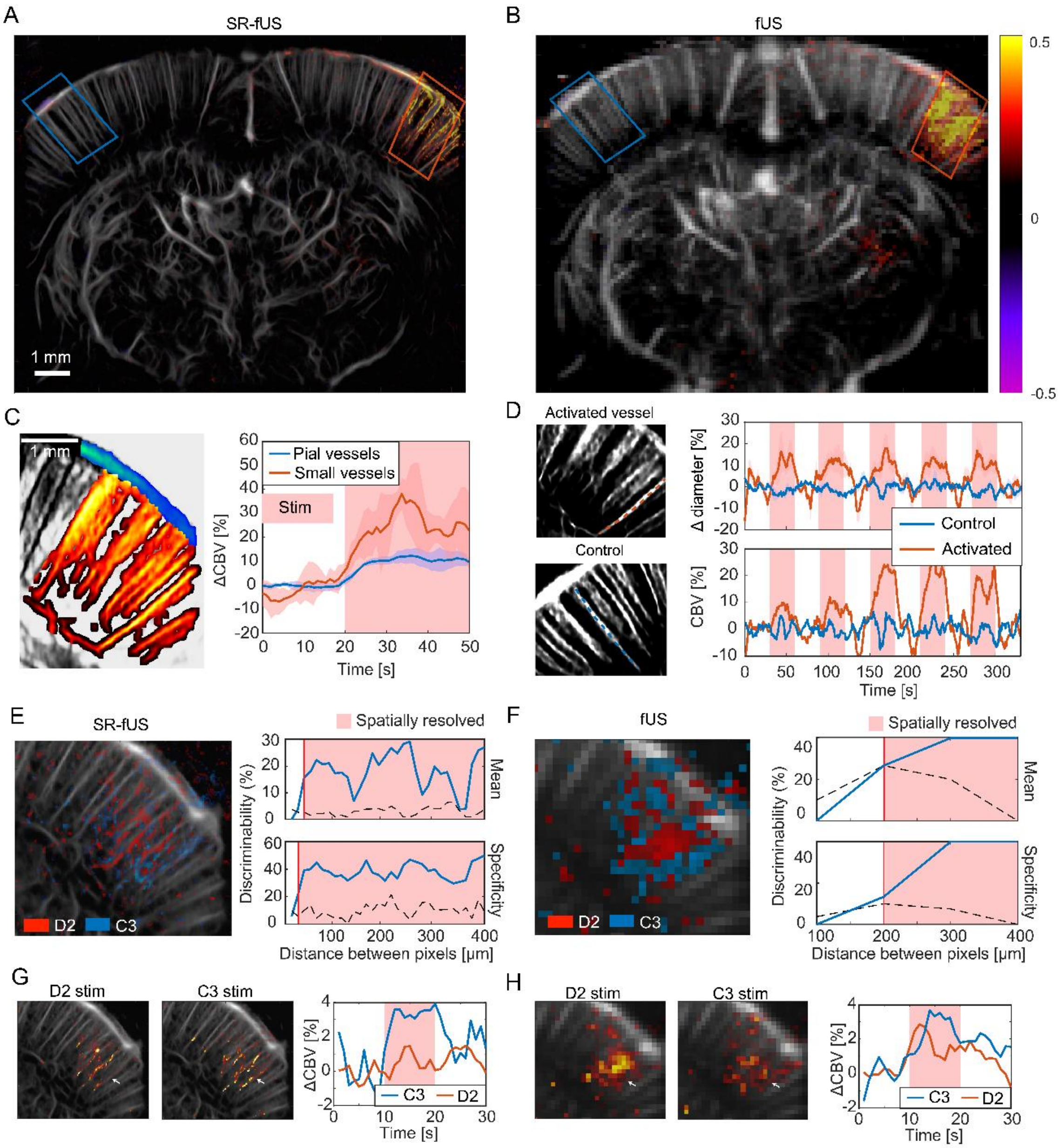

**Figure 3. SR-fUS revealed functional hyperemic changes in rat brain microvessels during whisker stimulation.** (**A** and **B**) fUS activation map and SR-fUS activation map during simple whisker stimulation. (**C**) Left: subdivision of the barrel cortex into pial vessels (blue) and small vessels (red) based on the super-resolution Doppler images. Right: mean CBV (± s.e.m.) of pial vessels and small vessels from n=5 courses. (**D**) Diameter and CBV changes of a single vessel marked by the red dashed line. (**E** and **F**) Left: activation maps generated by SR-fUS and fUS during single-whisker stimulation (D2, red; C3, green). Right: quantification of the lower spatial limit at which one can significantly find differences in the mean responsiveness (upper) or stimulus-specific responsiveness (lower) of two pixels, with respect to their distance. Dashed lines show the upper limit for chance (p<5e-2, 5\% percentile over 50 randomizations). (**G** and **H**) Left: CBV changes generated by SR fUS and fUS during stimulation of the D2 and C3, respectively. Right: CBV change curves for the single microvessel indicated by the white arrow.

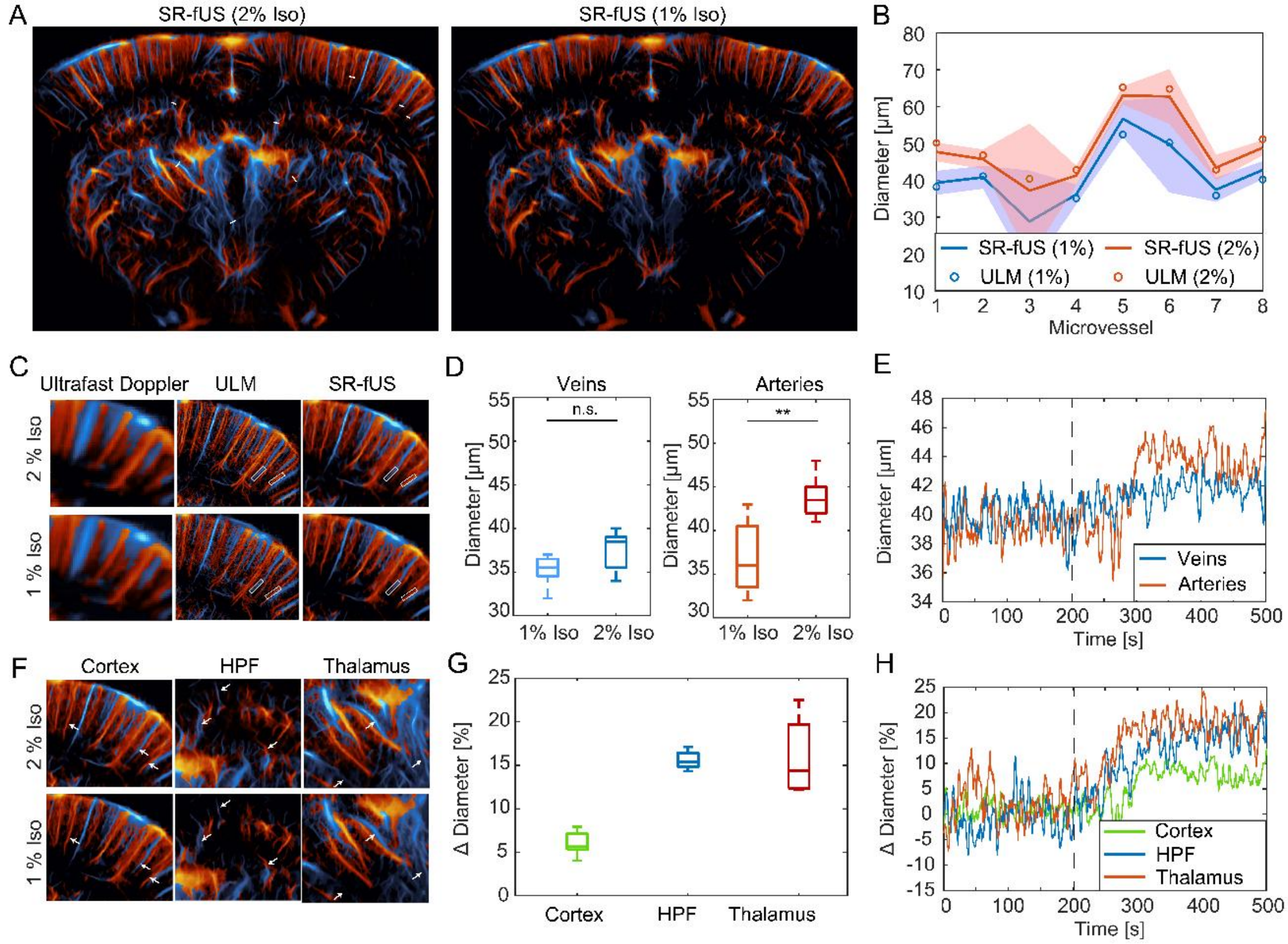

**Figure 4. SR-fUS revealed an increase in microvascular diameter induced by isoflurane anesthesia.** (**A**) SR-fUS images of rat brains under 2\% isoflurane anesthesia and 1\% isoflurane anesthesia (ISO: isoflurane). (**B**) Quantitative comparison of the microvessel size between the SR-fUS images and ULM images. Red and blue indicate mean diameter SR-fUS measurements (± s.e.m.) under 2\% and 1\% isoflurane anesthesia, respectively. N = 100, (1–100 s for 1\% isoflurane and 400–500 s for 2\% isoflurane). (**C**) Cortex regions of ultrafast Doppler, ULM and SR-fUS images. (**D**) Comparison of microvessel diameter measured by SR-fUS for the selected arterial and venous segments. Statistical analysis was conducted using t-test at each measurement point along the segments (white rectangles in C, take n=8 samples from each segment). (**E**) Dynamic curve of microvascular diameter. (**F**) SR-fUS images of the ROI regions in the cortex, hippocampal formation (HPF) and thalamus. (**G**) Relative changes in microvascular diameter across different regions. Microvessels are indicated by white arrows. Three diameter measurements were taken from different positions along each microvessel for statistical analysis. (**H**) Mean dynamic curves of microvascular diameter changes in different brain regions.

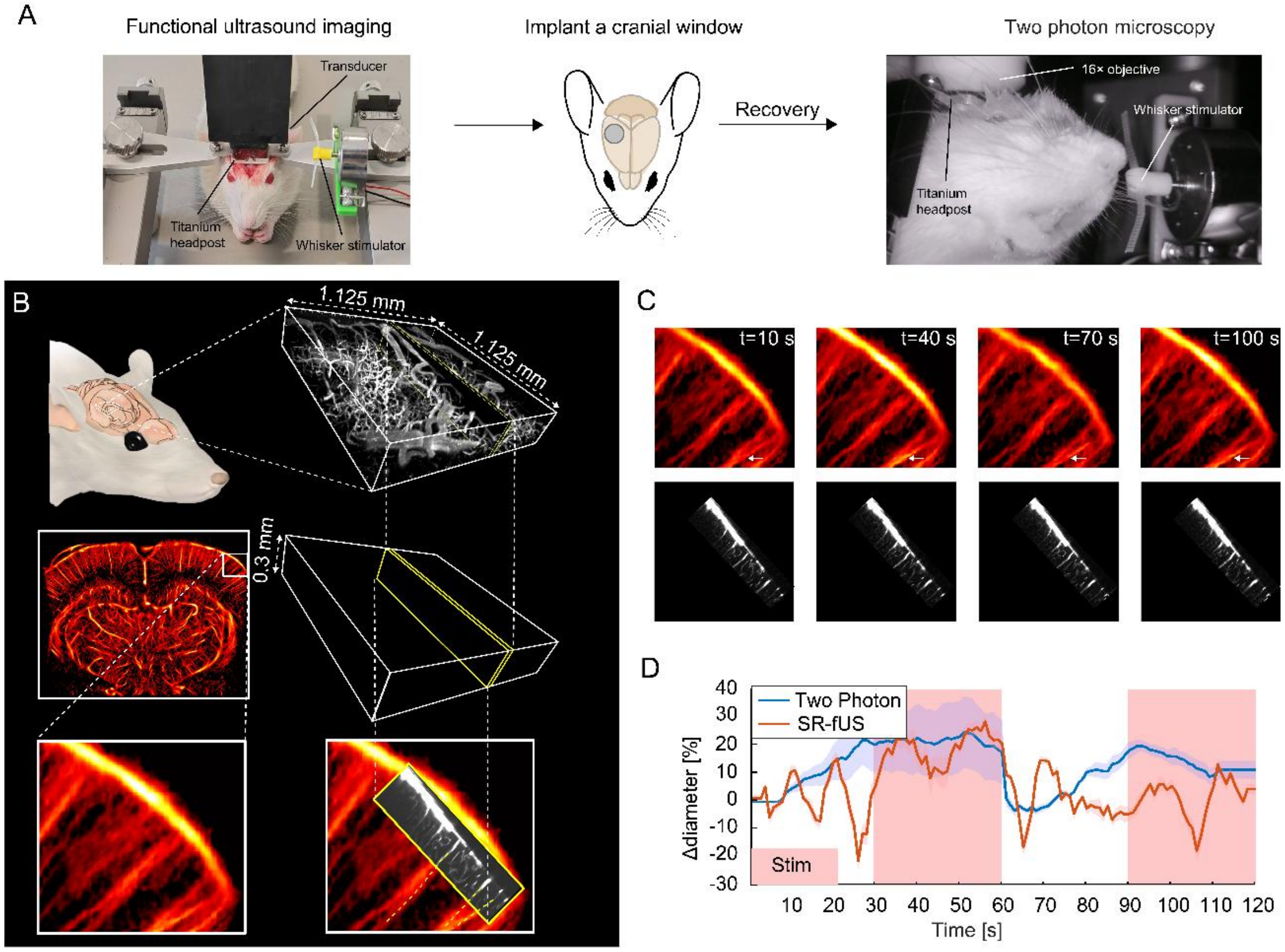

**Figure 5. Comparative results of SR-fUS and two-photon microscopy.** (**A**) Flowchart of a comparison experiment. Functional ultrasound imaging under whisker stimulation was first performed on awake craniotomized rats; then the rats were anesthetized and a cranial window was implanted in the corresponding brain region; two-photon microscopy imaging was performed after recovery from anesthesia. (**B**) Super-resolution Doppler images match well with two-photon microscopy images. (**C**) Representative images of SR-fUS and two-photon microscopy at different time points in whisker stimulation experiments. (**D**) Relative change curves of microvessel (shown as yellow dashed lines in the lower right of B) diameters obtained along the stimulation pattern based on SR-fUS and two-photon microscopy.

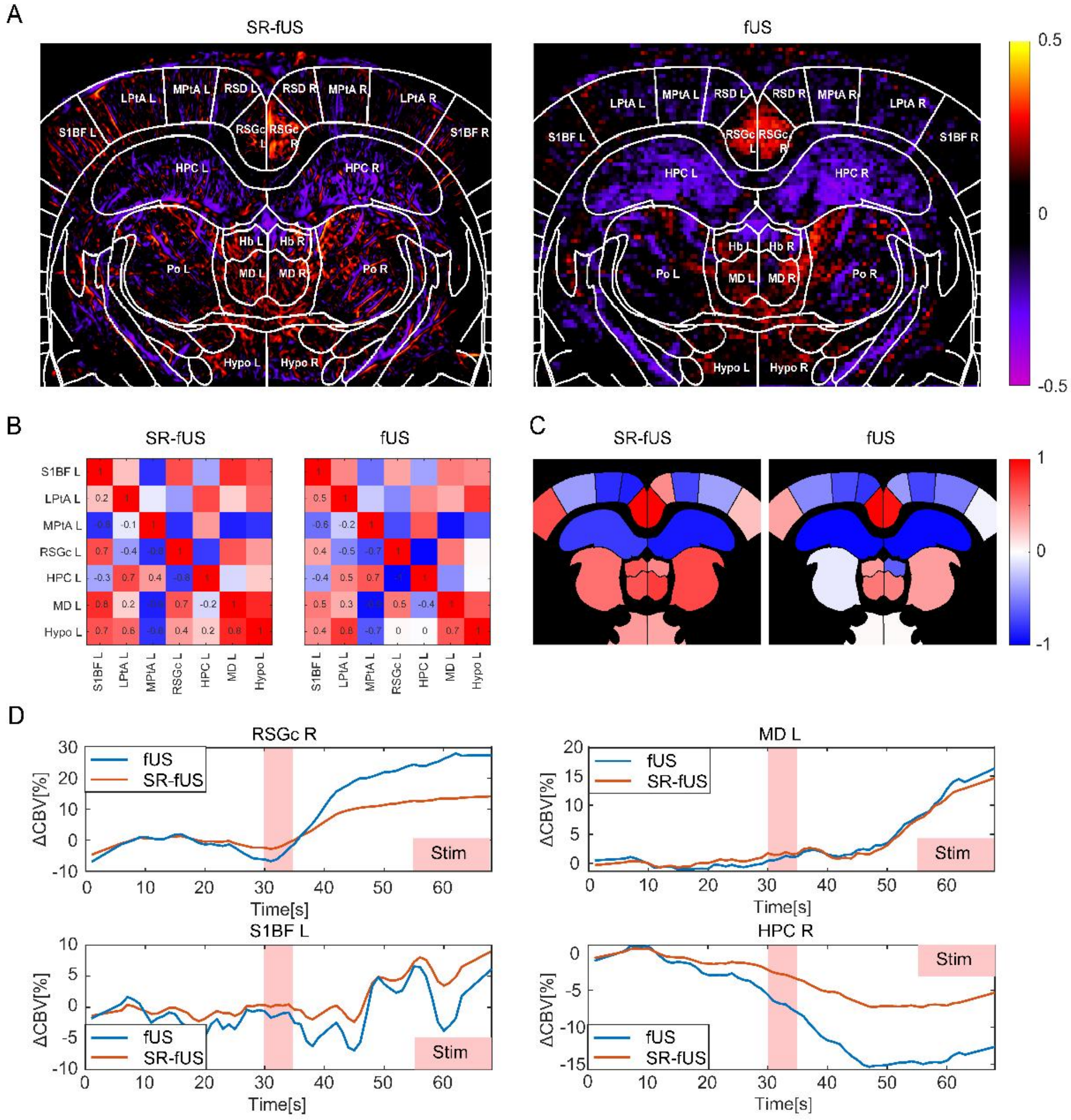

**Figure 6. SR-fUS reveals functional hyperemic responses to pain stimulation in rat brains.** (**A**) Correlation maps of CBV with stimulation patterns based on SR-fUS and fUS. (**B**) FC matrix based on SR-fUS and fUS in post-stimulus period. (**C**) Seed-based analysis of the FC based on SR-fUS and fUS in post-stimulus period, where the RSGc_L was selected as the seed region. Pixel values were determined by calculating the correlation coefficient between the mean CBV of each brain region and the seed region. (**D**) Comparison between CBV fluctuations obtained based on fUS and those obtained based on SR-fUS in RSGc_R, MD_L, S1BF_L and HPC_R.